\documentclass[fleqn,11pt]{SelfArx} 

\usepackage{lipsum} 
\usepackage{multirow}

\definecolor{color1}{RGB}{0,0,90} 
\definecolor{color2}{RGB}{0,20,20} 

\usepackage{hyperref} 
\hypersetup{colorlinks=true,urlcolor=blue,linkcolor=blue,linkbordercolor=blue,breaklinks=true,bookmarksopen=false,pdftitle={Title},pdfauthor={Author}}

\JournalInfo{Geomagnetism and Aeronomy, 2022, Vol. 62, No. 7, pp. 903--910.} 
\Archive{DOI: 10.1134/S0016793222070131} 
\Archive{}

\PaperTitle{
Activity of Selected Solar Twins
} 

\Authors{M.\,M.~Katsova\textsuperscript{a,*}, 
B.\,A.~Nizamov\textsuperscript{a},
and A.\,A.~Shlyapnikov\textsuperscript{b}
} 

\affiliation{\textsuperscript{a}\textit{
Sternberg State Astronomical Institute, Lomonosov Moscow State University, Moscow, Russia
}} 
\affiliation{\textsuperscript{b}\textit{
Crimean Astrophysical Observatory, Russian Academy of Sciences, Nauchny, Russia
}} 

\affiliation{\textsuperscript{*}e-mail: maria@sai.msu.ru, mkatsova@mail.ru}
\affiliation{Received: March 8, 2022; revised March 23, 2022; accepted May 5, 2022}

\Keywords{solar twins, stars: activity, stars: coronae, stars: rotation, stellar flares, lithium abundance}
\newcommand{\keywordname}{Keywords}

\Abstract{
We analyze various tracers of magnetic activity for 23 solar twins which 
are characterized by significant scatter of lithium abundance in their atmospheres. 
A level of coronal and chromospheric activity has been studied from 
available X-ray and UV-archival data. It gives us a chance to compare 
coronae of solar twins of various ages with the solar case. 
We found a scatter in the X-ray to bolometric luminosity ratio
 $L_{\rm X}/L_{\rm bol}$ by several orders of magnitude, which exists in these stars 
 along with a significant spread in Li abundance. 
This may link the surface activity of stars with phenomena at the base 
of their convective zones. The TESS data allowed us to reveal rotation 
modulation of stellar brightness associated with starspots. 
For some twins of our samples, periods of axial rotation are detected 
around 6 days, i.e. these stars rotate almost 4 times faster than the 
contemporary Sun. 
This indicates their relative youth. 
Flare activity of solar twins is discovered in the TESS data; 
we showed existence of various kinds of flares, and present temporal profiles for some of them. 
We obtained the energy about of $8 \times 10^{33}\,$erg for 
the largest flare of our samples, lasting longer than 4 h. 
In addition, we discuss also magnetic fields and exoplanets, 
orbiting these stars. 

~

\textbf{DOI} ~ 10.1134/S0016793222070131
}

\begin{document}

\flushbottom 

\maketitle 


\thispagestyle{empty} 

\section{INTRODUCTION}

The solar-like magnetic activity phenomena are inherent in stars 
on the lower part of the main sequence. 
Among solar-type stars there are solar analogues with fundamental 
parameters similar to the solar ones, and the solar twins can be 
identified as stars which have the same physical properties as those 
of the Sun: mass, radius, luminosity, chemical composition, rotation, and activity. 
This should mean that the spectrum of a solar twin should be identical to that of the Sun. 
However there is a set of solar twins that have a significant scatter
 in the lithium abundance in their atmospheres (Mishenina et al., 2020). 
Understanding the reasons for this behaviour of lithium is important in 
studying the stellar evolution and the chemical evolution of the Galaxy. 
Insofar as $^7$Li is destroyed at temperatures $\sim 2.5 \times 10^6\,$K, 
knowledge of the abundance of lithium provides an effective tool for 
studying the various mixing processes inside solar-type stars that lead 
to changes in its abundance. 
These processes include, first of all, convection, microscopic diffusion, 
meridional circulation, and chromospheric activity. 
In addition to the initial conditions with which a star enters the main 
sequence, one of the factors affecting the abundance of lithium may be 
the activity (Mishenina et al., 2012). 
The latter is important point in understanding the evolution of solar-type 
activity during the further life of a star on the main sequence (Katsova et al., 2013).

\begin{figure*}[!th] 
\centering
\includegraphics[width=\linewidth]{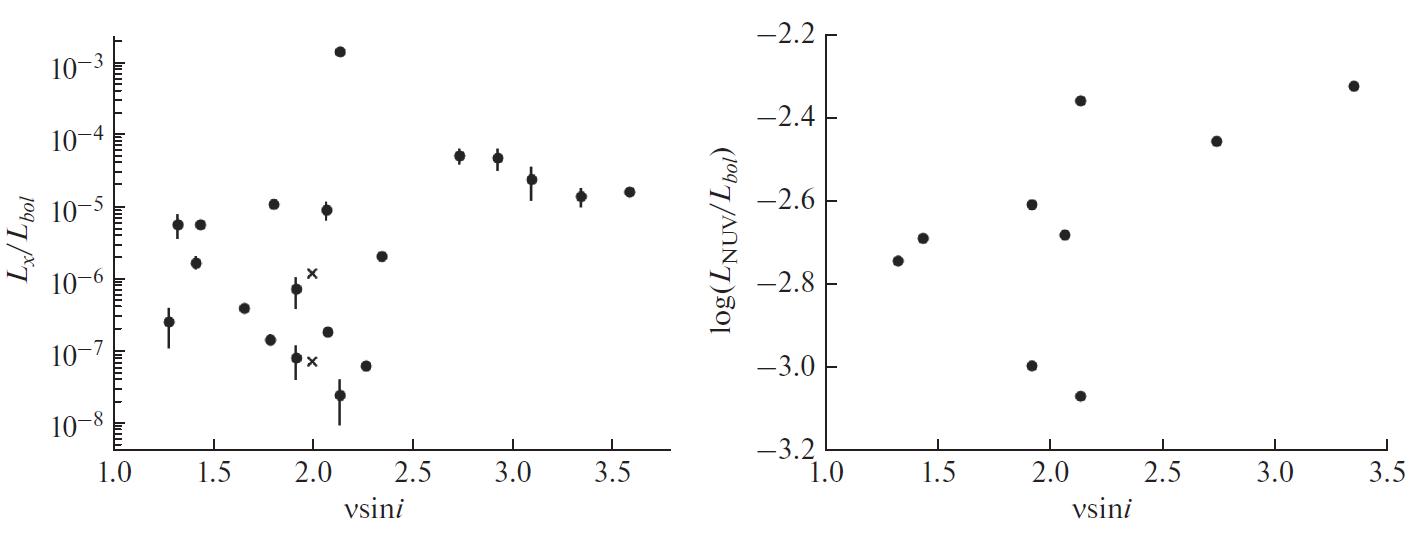}
\caption{
The left panel: the coronal activity index, $L_{\rm X}/L_{\rm bol}$ for solar twins 
versus rotation velocities $v \sin i$; two values for the Sun at minimum 
and maximum activity level are also plotted as crosses (Peres et al., 2000). 
The right panel: the near UV- to the bolometric luminosity ratio versus 
the velocities $v \sin i$.
}
\label{Figure1}
\end{figure*}

We consider here evidences for activity of 23 solar twins out of 55 solar analogues 
for which effective temperature, gravity $\log g$, and metallicity [Fe/H] differ from 
the solar values by maximum 100 K, 0.1 dex, and 0.1 dex respectively 
(Mahdi et al., 2016). Mishenina et al. (2020) found significant 
scatter of lithium abundance in their atmospheres. 
These targets are proxies of the Sun that essentially differ only in their rotation 
periods (and, therefore, age) and hence in the efficiency of the magnetic dynamo. 
For a given main-sequence stellar mass, the rotation period is a rather 
well defined function of the star's age if the latter exceeds a few 100 Myr. 
Magnetic activity expresses itself in the outer stellar atmospheres 
of these stars through various phenomena such as dark magnetic spots, 
bright chromospheric plages, chromospheric emission lines, and coronal 
radio and X-ray emissions. These phenomena are characterized by several 
key tracers such as axial rotation, X-ray and UV-radiation, magnetic 
fields, and non-stationary events as well. In particular, now it became 
clear that the coronal emission displays the largest range of variation 
depending on the surface magnetic activity level. The saturated regime 
of activity changes to the unsaturated mode, when the solar-type activity 
is formed. We analyze here relevant characteristics of these stars, compare 
them with the contemporary Sun and discuss them in the context of an 
evolution of solar-stellar activity. It is obviously that both X-rays 
and UV-radiation from the stellar coronae can affect the planet 
atmospheres, therefore this study is significant for stars with 
various activity levels.

\section{THE X-RAY AND UV-EMISSION OF SOLAR TWINS WITH THE DETECTED LITHIUM}

We analyzed available data for solar twins with detected lithium and found 23 stars out of 55 solar analogues, for which the X-ray emission was registered during such space missions as Einstein, EXOSAT, ROSAT, XMM-Newton and Chandra. Besides, definite information about UV-radiation of these stars can be derived from the GALEX archival data. Because both kinds of high energy radiation determine properties of outer atmospheres of low mass stars, they allowed us to judge about a level of their activity. 

The general parameters of these stars are given in Table 1 together with data about Li abundances. The coronal index, the X-ray to bolometric luminosity ratio, $R_{\rm X} = L_{\rm X}/L_{\rm bol}$, we estimated for several samples of solar twins from the above mentioned X-ray archival data (Evans et al., 2010; Boller et al., 2016; Traulsen et al., 2020; the corresponding references are given in the description to Table 1). Dependence of the obtained coronal indices on rotational velocities is presented in Fig. 1 (the left panel).The corresponding values for the Sun at minimum and maximum activity level are also plotted for comparison by crosses (after Peres et al., 2000). In spite of spread of our values, definite trend of increase of the X-ray luminosities for stars with the faster rotation does exist.

\begin{table*}[!p] 
\caption{
\bf
\ \ Physical parameters of the stars 
}
\par\medskip
\begin{center}
\begin{tabular}{r|c|c|c|c|c|c|c|c}
\hline
\multirow{3}{1.2cm}{~ ~ HD} & 
\multirow{3}{1cm}{~~$d$, pc} & 
\multirow{3}{1cm}{$\log A_{\rm Li}$} & 
\multirow{3}{1.1cm}{~~$v \sin i$, \mbox{\ \ km/s}} & 
\multirow{3}{1.5cm}{$L_{\rm bol}/L_{\rm Sun}$} & 
\multirow{3}{1.4cm}{~~~~$L_{\rm X}$, $10^{27}\:$erg/s} & 
\multirow{3}{1.8cm}{$\log L_{\rm X}/L_{\rm bol}$} & 
\multirow{3}{2.1cm}{$\log L_{\rm FUV}/L_{\rm bol}$} & 
\multirow{3}{2.1cm}{$\log L_{\rm NUV}/L_{\rm bol}$} \\
&&&&&&&&\\
&&&&&&&&\\
\hline
5 294   & 30 & 2.17  &  3.35  &  0.84  &  43    &  --4.861  &  --4.9   &  --2.3  \\
9 986   & 25 & 1.75  &  2.07  &  1.10  &  37    &  --5.051  &  --5.1  &  --2.7  \\
24 409  & 23 & 1.15  &  1.92  &  1.11  &  3.0   &  --6.143  &  --5.1  &  --2.6  \\
29 150  & 34 & 1*    &  ---   &  0.93  &  13    &  --5.421  &  ---     &  ---  \\
42 618  & 24 & 1.3   &  1.28  &  0.94  &  0.9   &  --6.604  &  ---     &  ---  \\
42 807  & 19 & 2.1   &  3.59  &  0.81  &  49    &  --4.801  &  ---     &  ---  \\
56 124  & 27 & 2.0   &  1.33  &  1.07  &  23    &  --5.251  &  --5.0  &  --2.7  \\
75 767  & 23 & 1.45  &  3.1   &  1.07  &  97    &  --4.622  &  ---     &  ---  \\
76 151  & 17 & 1.75  &  1.44  &  0.97  &  21    &  --5.254  &  --5.0  &  --2.7  \\
95 128  & 14 & 1.8   &  2.14  &  1.58  &  0.15  &  --7.615  &  ---     &  --3.1  \\
117 176 & 18 & 1.78  &  ---   &  3.04  &  0.93  &  --7.095  &  --5.5  &  --3.4  \\
140 538 & 15 & 1.5   &  ---   &  0.86  &  25    &  --5.121  &  --5.2  &  --2.7  \\
146 233 & 14 & 1.6   &  1.79  &  1.09  &  0.60  &  --6.844  &  ---     &  ---  \\
159 222 & 24 & 2.0   &  2.74  &  1.24  &  235   &  --4.301  &  --5.1  &  --2.5  \\
168 009 & 23 & 1*    &  1.92  &  1.41  &  0.43  &  --7.106  &  --5.1  &  --3.0  \\
181 655 & 25 & 1.93  &  1.81  &  1.72  &  71    &  --4.961  &  ---     &  ---  \\
186 408 & 21 & 1.45  &  2.08  &  1.57  &  1.1   &  --6.749  &  ---     &  ---  \\
186 427 & 21 & 0.8*  &  2.27  &  1.25  &  0.30  &  --7.205  &  ---     &  ---  \\
187 237 & 26 & 2.15  &  2.35  &  1.01  &  7.8   &  --5.694  &  ---     &  ---  \\
222 143 & 23 & 2.0   &  2.93  &  1.05  &  186   &  --4.332  &  ---     &  ---  \\
224 465 & 24 & 0.8*  &  1.42  &  1.02  &  6.5   &  --5.777  &  ---     &  ---  \\
146 362 & 23 & 2.43  &  2.14  &  1.02  &  5300  &  --2.864  &  ---     &  --2.4  \\
187 123 & 46 & 1.2   &  1.66  &  1.34  &  2.0   &  --6.48   & ---      &  ---  \\
\hline
\end{tabular}
\end{center}

\begin{flushleft}
The following data are given: Henry Draper catalog id, distance, lithium abundance 
(asterisks denote upper limits), rotation velocity projection, bolometric luminosity 
in solar units, logarithms of the ratios of the X-ray, far-ultraviolet (1350--1750 \AA) 
and near-ultraviolet (1750--2750 \AA) luminosities to the bolometric one. 
Distances are taken from the SIMBAD database, lithium abundances from the work by 
Mishenina et al. (2020), bolometric luminosities from the TIC catalog, 
UV luminosities from GALEX. The sources of the X-ray luminosity are denoted by 
upper numbers as follows: \\
$^1$Second ROSAT all-sky survey (2RXS) source catalog 
(Boller+, 2016) 2016A\&A...588A.103, \\
$^2$XMM-Newton slew survey Source Catalogue, version 2.0 (XMM-SSC, 2017), \\
$^3$ROSAT All-Sky Survey Faint Source Catalog (Voges+ 2000), IAU Circ., No. 7432, \#3 (2000), \\
$^4$XMM-Newton Serendipitous Source Catalogue 4XMM-DR9 (Webb+, 2020) 2020A\&A...641A.136, \\
$^5$Chandra observations of solar analogs (Miller+, 2015) 2015ApJ...799..163, \\
$^6$The Chandra Source Catalog (CSC), Release 2.0. Evans, et al., 2019, HEAD, 114.01, \\
$^7$The WGACAT version of ROSAT sources (White+ 2000) \url{https://heasarc.gsfc.nasa.gov/wgacat/}, \\
$^8$Kashyap et al. 2008 ApJ 687, 1339, \\
$^9$Classification of Swift and XMM-Newton sources (Tranin+2022) 2022A\&A...657A.138.
\end{flushleft}

\label{Table1}
\end{table*}

This agrees with well known rotation-activity relation, indicated that 
late-type stars (G to M) have X-ray luminosities strongly dependent on rotation rate 
$L_{\rm X} \sim (v \sin i)^{1.9 \pm 0.5}$ and independent on bolometric luminosity (Pallavicini et al., 1981). 

The UV-radiation characterizes activity levels of the chromosphere and 
the chromosphere-corona transition region. Therefore we estimated also 
the UV-radiation for our stars in two spectral ranges: far-ultraviolet 
(FUV: 1350--1750$\:$\AA) and near-ultraviolet (NUV: 1750--2750$\:$\AA) 
bands observed with GALEX (Bianchi et al., 2011). 
The FUV-and NUV-luminosities are derived by us for several stars. 
We obtained that the minimal value of the FUV-luminosity differs from the maximal one by a factor of 1.8, 
while the similar difference for $L_{\rm NUV}$ reaches 4. 
The ratio of both parameters, $L_{\rm NUV}$ and $L_{\rm FUV}$ to the bolometric luminosities, are listed in Table 1. 
The available values of $\log L_{\rm NUV}/L_{\rm bol}$ are presented in Fig. 1 (the right panel). 
Although there is scatter in data, one can trace that the faster the rotation, the larger the UV-luminosity.

\begin{figure}[!t] 
\centering
\includegraphics[width=\linewidth]{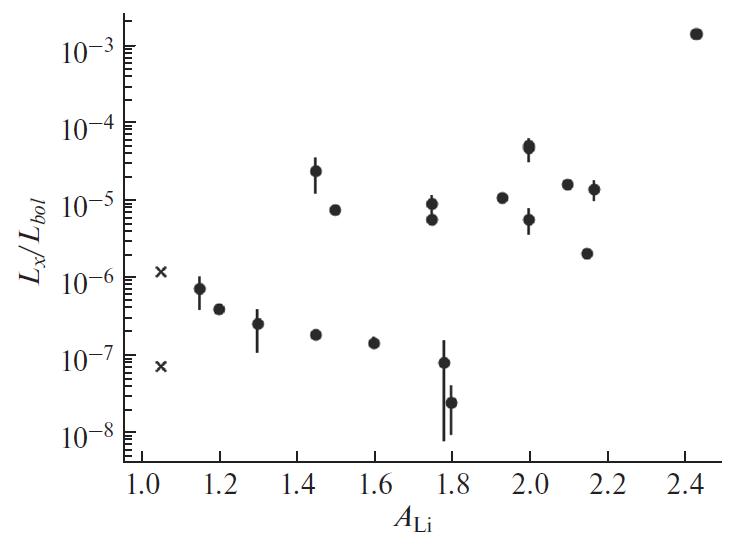}
\caption{
The coronal index versus lithium abundances for
solar twins considered. The solar Li content A(Li) = 1.05
is adopted from Asplund et al. (2009).
}
\label{Figure2}
\end{figure}

Thus, we found that the X-ray luminosities of the solar twins cover 
the wide range of values from the saturation limit in young stars, 
$R_{\rm X}$ around $10^{-3}$, to levels approximately 4 orders of magnitude lower. 
This significant scatter may mean that among the solar twins exist 
stars both more active and even less active than the Sun. 

Analyzing the activity of stars, in which the abundance of lithium 
shows a scatter of almost an order of magnitude, we note that among 
our twins there are objects of different ages from 1.5 to 5 Gyr 
(Fig. 4 in Mishenina et al., 2020). Further comparison of the 
coronal X-ray emission with lithium abundances for stars of our 
samples and the Sun shows that despite the spread, a certain trend 
can be traced, indicating a relationship between the coronal indices, 
the Li content and stellar ages. 
Indeed, Fig. 2 demonstrates that the higher coronal indices correspond in general to larger 
values of the Li content. This reflects the dependence of the activity level on stellar age. 
Thus, among the solar twins, we have identified stars with a more 
powerful corona, i.e. more active and relatively younger stars, 
apparently rotating a few times faster than the nowadays Sun. 
At the same time, there are solar twins, which have a lower level 
of activity than that of the contemporary Sun; their soft X-ray 
radiation is almost an order of magnitude weaker than the solar value.

\section{THE TESS BRIGHTNESS VARIABILITY OF SOLAR TWINS}

\subsection{Rotational Modulation}

The data from the TESS observatory, after preprocessing and filtering, 
make it possible to obtain the optical light curves for the considered 
stars (Ricker et al., 2014). We present as preliminary results a few 
examples of rotational modulation associated with spots. 
So, in the upper panel in Fig. 3 the TESS light curves for two solar 
twins HD 5294 and HD 24409 is 
shown. It is seen that HD 5294 demonstrates the higher amplitude of 
rotational modulation as compared with HD 24409 where both the level 
of the flux and amplitude variations are significantly lower. Thus, 
HD 5294 can be considered as more active star. This point is supported 
by comparison of their coronal indices which differ by more than one 
order of magnitudes. Note also that HD 5294 has twice higher Li 
abundance relatively to the solar value.

\begin{figure*}[!th] 
\centering
\includegraphics[width=\linewidth]{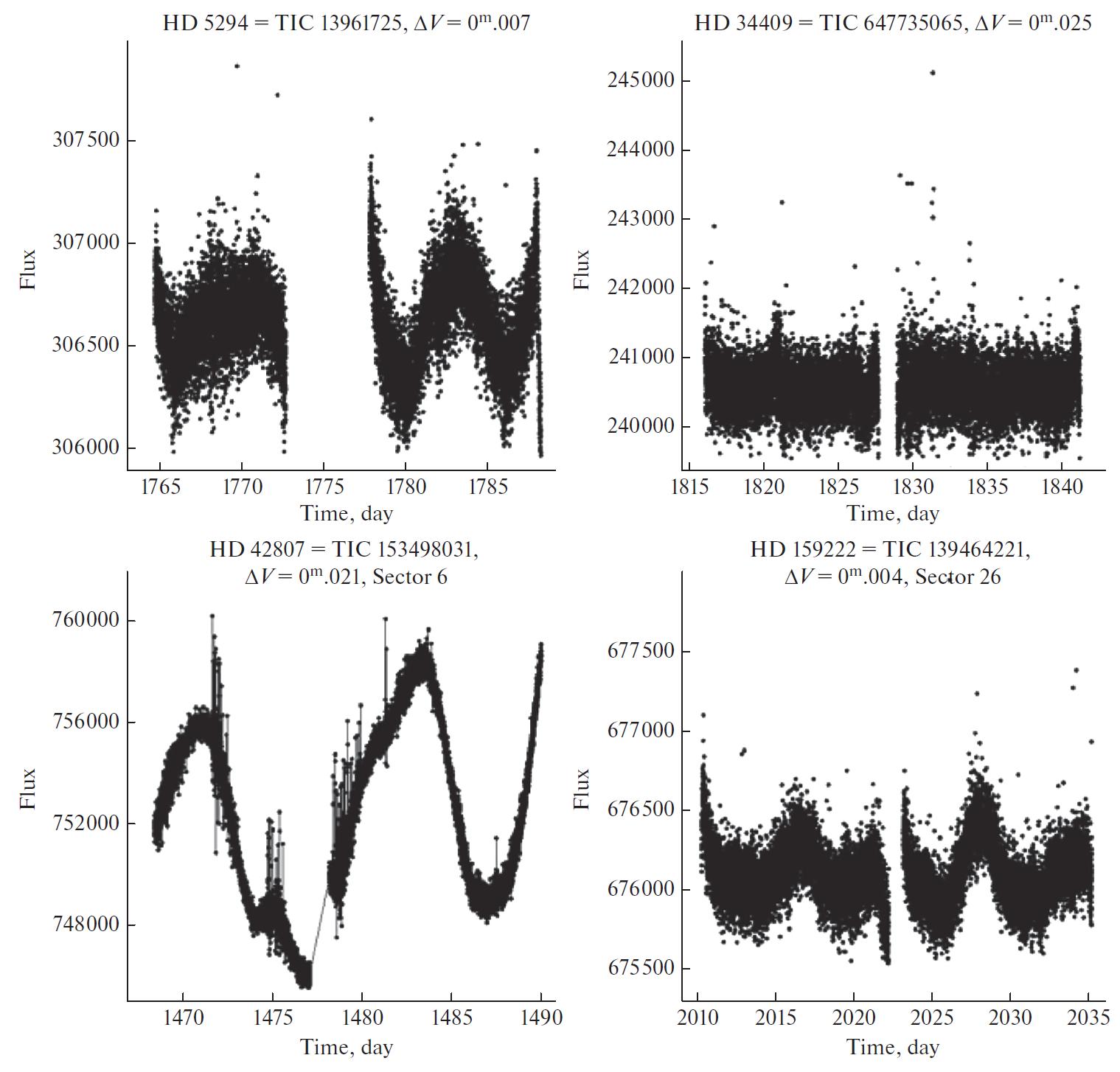}
\caption{
Samples of the TESS light curves for several solar twins. The upper panel: rotational modulation of HD 5294 and
HD 22409. The lower panel -- brightness variations of HD 42807, which is known as a chromospherically active RS CVn-type
star (the left graph), and a component of spectral binary HD 159222 (on the right).
}
\label{Figure3}
\end{figure*}

The light curve in Fig. 3 (left in the lower panel) for HD 42807 
presents a more complicated shape of the brightness variation with 
presence of flare events. This star has twice higher flux level than 
that from HD 5294, more expressed rotational modulation, and similar 
coronal indices, rotational velocity and lithium abundance as well. 
It should be noted here that HD 42807 is known as a chromospherically 
active RS CVn-type binary. The bottom right panel in Fig. 3 shows the 
TESS light curve for HD 159222 where the flux is also by a factor 2 
more than that for HD 5294, and activity parameters are similar to 
those in other stars in this Fig. 3. It cannot be excluded that 
this star can be a spectroscopic binary.

We estimated the period of axial rotation for two stars HD 5294 
and HD 159222 using a discrete Fourier transform algorithm 
(Lenz and Breger, 2014). We obtained the axial rotation period 
for HD 5294 is 6$^{\rm d}$.526949 for this set of observations. The accuracy 
of the frequency from which the period was found is 0$^{\rm d}$.011368. 
HD 159222 rotates with the period of 5$^{\rm d}$.92057. The accuracy of the 
frequency from which the period was found is 0$^{\rm d}$.008882. Results of 
convolution with found periods for both stars are presented in 
Figs. 4 and 5. Note that possible rotation periods for HD 5294 and 
HD 159222 were given also in the catalogue by Oelkers et al. (2018). 
However, period 0$^{\rm d}$.920615 for HD 5294 is found there in 47 more stars, 
and period 29$^{\rm d}$.533400 for HD 159222 is found in 792 stars. 
This calls into question the reliability of these periods. 
Convolutions with these periods do not show a real result. 
Thus, these quite active solar twins rotate with periods of 
around 6 days, i.e. 4 times faster than the contemporary Sun. 
This fact as well as their other tracers of activity indicates their comparative youth.

\begin{figure*}[!th] 
\centering
\includegraphics[width=\linewidth]{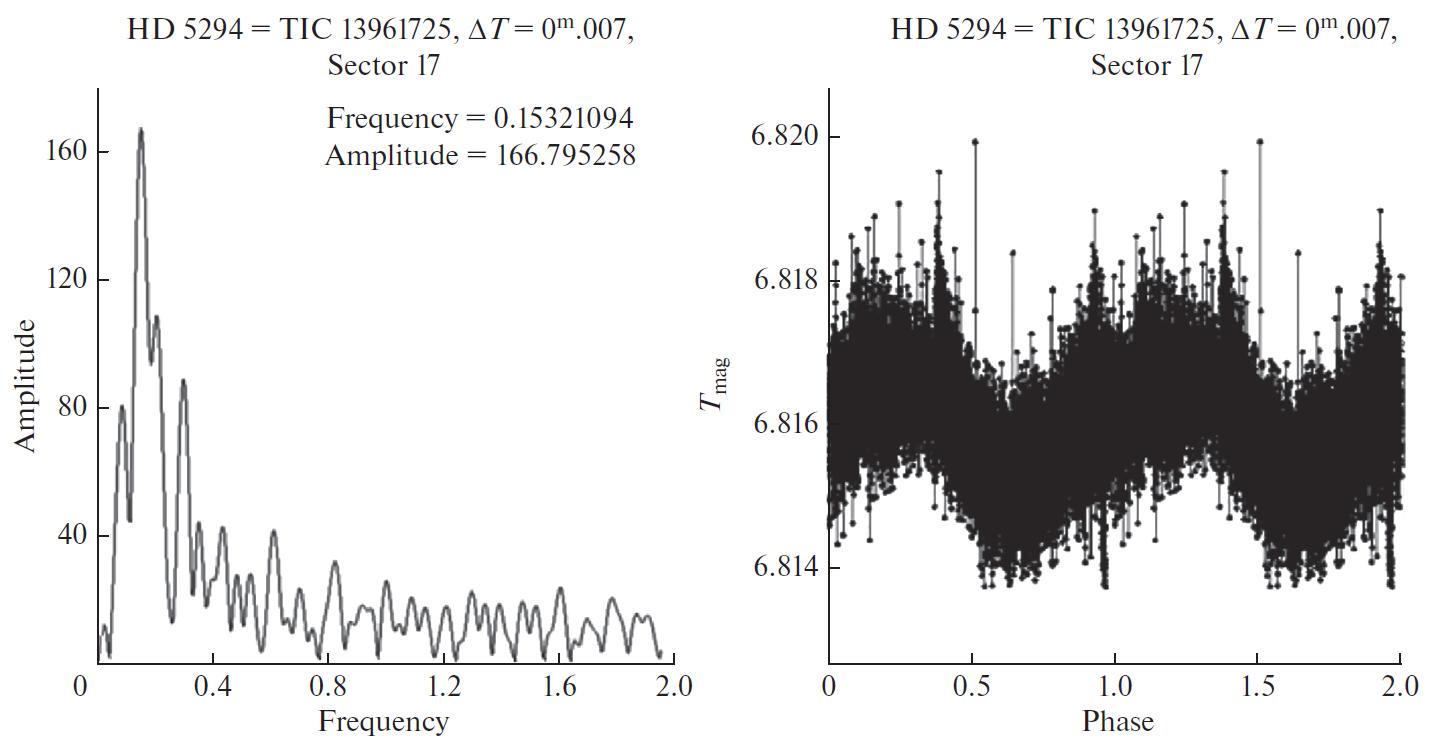}
\caption{
Frequency analysis diagram of time series for the HD 5294 star 
(constructed from the TESS data obtained in sector 17) on the left panel. 
Convolution with a period of 6$^{\rm d}$.526949 on the right panel. 
The accuracy of the frequency from which the period was found is 0$^{\rm d}$.011368. 
}
\label{Figure4}
\end{figure*}

\subsection{Optical Flares on Solar Twins Registered with the TESS Observatory}

Considering flare activity of solar analogues, we have to keep in mind 
that the most powerful flares on the Sun are rare phenomena. Flares in 
optical continuum emission which are analogues of solar white-light 
flares have not been seen on single main sequence G stars until the 
Kepler mission. Their detection both on the Sun and on these stars 
is difficult because a source of the flare optical continuum emission 
has a low contrast against the photosphere, a small flare area, and 
lives a short time (a few minutes). Nevertheless, an existence of 
optical flares in G stars can be suspected in the UV data supports by 
recent study by Kashapova et al. (2021) where was shown that time 
profiles of solar flares in the UV-continuum emission are similar 
to impulsive flare light curves registered on the red dwarf during 
the Kepler mission. Indeed, flares on G stars were found in 
GALEX NUV-data (Brasseur et al., 2019). 

The only recently a definite information appears about flare activity 
of these stars from other observations, including ground-based monitoring 
(Jackman et al., 2018; Bondar' et al., 2021; Koller et al., 2021). 
However, until now these surveys do not contain data for solar twins of our sample. 

The first Kepler data about flares on G dwarfs were reported by 
Maehara et al. (2012, 2015). Later on it was found that the energy of 
some of flares can reach a few times of $10^{36}\,$erg (e.g. Wu et al., 2017). 

After completion the Kepler mission, the TESS space observatory 
provides possibility to continue permanent monitoring of larger 
number of stars that allows searching for flares on these stars. 
We have analyzed TESS light curves for some of our stars. 
We selected several high amplitude, non-single-impulse events, using 
the $3\sigma$-criterion and give some samples of TESS flare light curves. 
The temporal profiles for 3 flare events on the solar twins HD 24409 
and HD 75767 registered with time resolution 120 s are presented in Fig. 6a. 
Note the quite complicated shape of the flare light curve in the 
middle of this figure. The third flare sample on the HD 75767 star 
looks like a classical impulsive flare on red dwarfs, but it occurs 
on the G star. We have been estimated the energy of these non-stationary 
events. For the HD 24409 star, the lower limit of the total energy of 
the larger flare (in the middle of the left panel) is 
$8 \times 10^{33}\,$erg, its duration is 
4.32 h. The two flares to the right and left of it are about of 2 times 
weaker in the energy, around $4 \times 10^{33}\,$erg, and last 0.5--1 h less. 
As for the flares on the HD 75767 star, the flare, demonstrating quite 
complicated light curve shape, has the energy $4.12 \times 10^{33}\,$erg and 
its duration is 12 h, while the next sample of flares on the left panel 
in Fig. 6 looks like a classical impulsive flare on red dwarfs, but it 
occurs on the G star. This flare of the larger amplitude with the 
total energy of $4.3 \times 10^{33}\,$erg lasts 21.6 h.

\begin{figure*}[!t] 
\centering
\includegraphics[width=\linewidth]{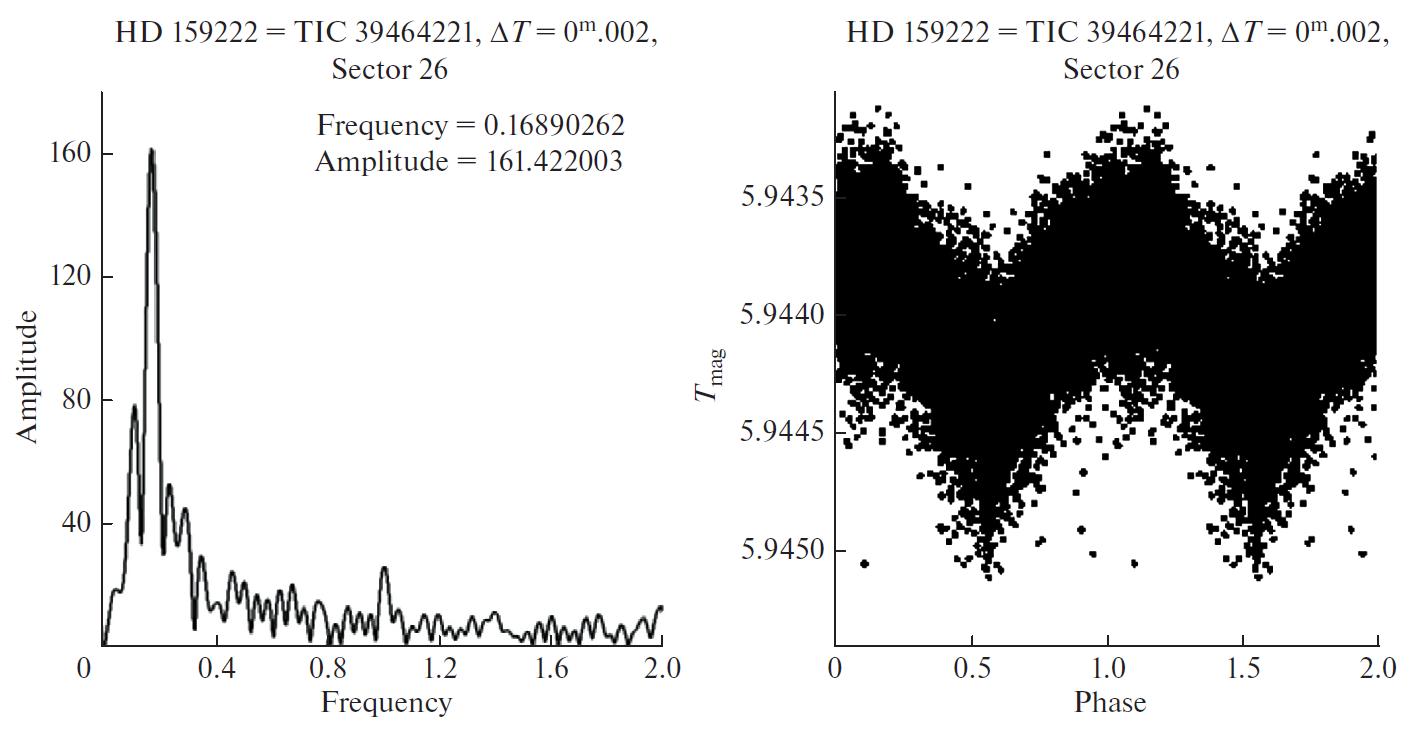}
\caption{
Frequency analysis diagram of time series for the HD 159222 star 
(constructed from the TESS data obtained in sector 26) on the left 
panel. Convolution with a period of 5$^{\rm d}$.9205 on the right panel. 
The accuracy of the frequency from which the period was found is 
0$^{\rm d}$.008882. 
}
\label{Figure5}
\end{figure*}

\section{ON THE MAGNETIC FIELD OF SOLAR TWINS}

There is reason to believe that activity of solar analogues at various 
levels from the highest one to low activity of the contemporary Sun is 
caused by local magnetic fields with a definite governing role of the 
large-scale magnetic field. Therefore, analyzing the factors determining 
activity, it needs to know information about stellar magnetic fields. 
Earlier, we compared data for magnetic fields obtained in the frameworks 
of the "Bcool collaboration" by Marsden et al. (2014) for several 
active G stars with data on the general
magnetic field of the Sun as a star. We obtained
the average value of the modulus of the magnetic field
$|B_l| = 4.72 \pm 0.53\,$G for G stars (Katsova and Livshits, 2014). 
The obtained average value corresponds to a 
rotation rate of about 4 km/s. Thus, the mean value of
the longitudinal fields for active G stars, rotating
slower than 5 days, is of one order of magnitudes
stronger than the mean daily magnetic field of the Sun as a star, 
which does not exceed 0.5 G near the maximum of the cycle. 
Now we considered these data for the solar twins, and it turned 
out that for only a few stars measurements can be considered reliable. 

We chose value of mean surface longitudinal magnetic field, $B_l$, 
exceeding $3\sigma$ from Table 3 of Marsden et al. (2014), which 
are marked as a "definite detection". The only 5 of 
our 23 solar twins possess observable magnetic fields. These stars 
with magnetic fields are HD 9986: $B_l = +2.5 \pm 0.5\;$G; 
HD 56124: $B_l = +4.9 \pm 0.8\;$G, HD 76151: $B_l = -3.7 \pm 0.2\;$G, 
HD 146233: $B_l = -2.3 \pm 0.4\;$G, and HD 222143: $B_l = -5.2 \pm 0.8\;$G. 
Note that all these stars differ from other solar twins in two 
significant factors. Firstly, they possess an enhanced Li abundance 
relatively to the Sun: $\log A_{\rm Li}$ from 1.6 to 2.0. The second 
difference is that the radius of the convective zones of these stars 
in the range $0.260-0.289\;R_{\rm Sun}$, while estimates for the rest 
twins are from 0.325 to $0.371\;R_{\rm Sun}$, as it follows from 
Table 1 in Marsden et al. (2014). Both circumstances point to relative 
youth of these stars as compared with other solar twins.

\begin{figure*}[!t] 
\centering
\includegraphics[width=\linewidth]{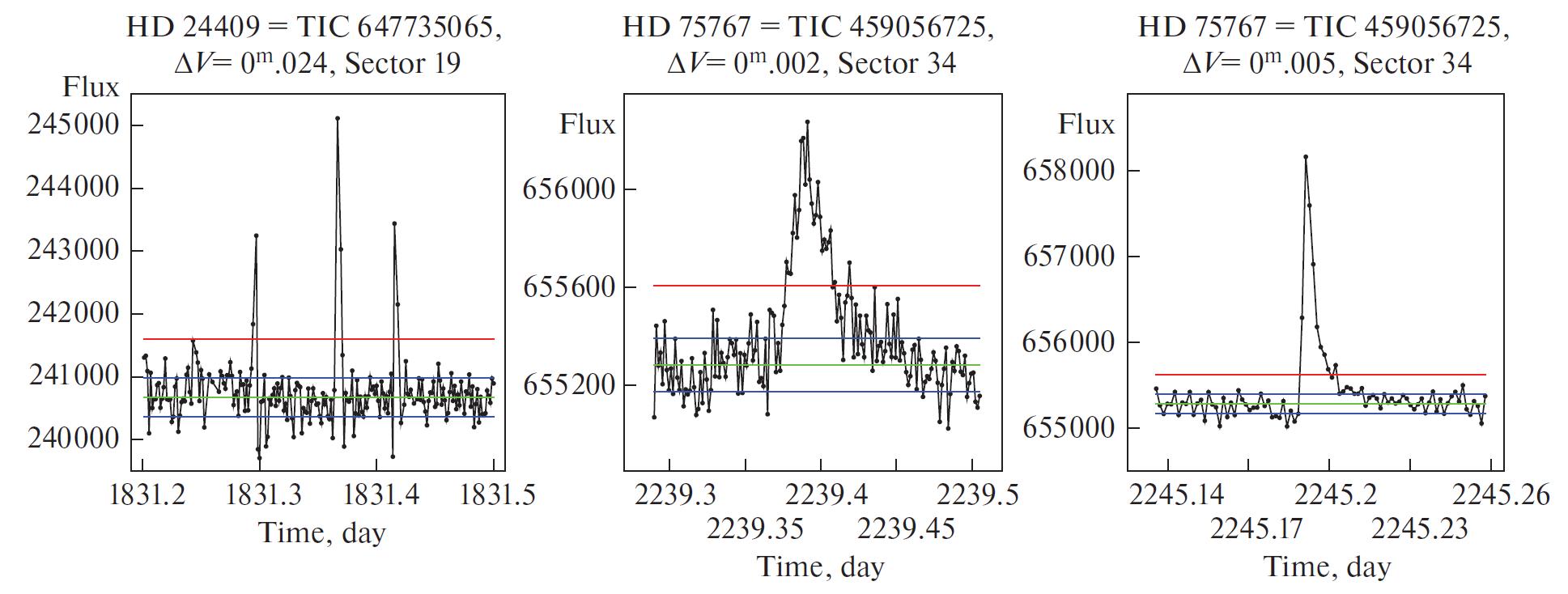}
\caption{
The TESS light curves for one of the flareson the HD 24409 star (the left panel) 
and two flareson HD 75767 registered with time-resolution 120 s. 
The green line corresponds to the mean level of stellar brightness in quiescence; 
two blue lines indicate the $\pm 1\sigma$-variability level, and red line is the 
$3 \sigma$-level. 
}
\label{Figure6}
\end{figure*}

\section{DISCUSSION AND CONCLUSION}

Thus, our analysis of various tracers of stellar activity for solar twins with detected lithium abundances shows that this group of twins contains definitely stars of various ages. It follows from a significant scatter in the coronal indices, which are close to the saturation activity level for the most active young dwarfs. This spread reflects similar dispersion in the lithium content. Besides, magnetic field measurements showed that mean longitudinal magnetic fields can be 5--10 times stronger as compared with the nowadays solar 
value, that also indicates that these stars are younger that the others. Such magnetic fields can be able, in principle, to provide the energy for powerful flares, much stronger than the solar ones. Studying flare activity and other powerful non-stationary phenomena is of importance for atmospheres of exoplanets orbiting similar stars. 
Note that because ages and mass of solar twin are defining in lithium abundance (do Nascimento et al., 2009), non-standard mixing is required to explain the low Li abundances observed in solar twins. So, for example, the 16 Cygni system consists of two stars HD 186408 (16 Cyg A is a good candidate for the most Sun-like star) and HD 186 427 having the same age, the same initial composition, and the Li abundance differences currently observed must be due to their various paths of evolution, related to the fact that one of them hosts a planet while the other does not. 
By the way, note here that 6 solar twins have exoplanets orbiting these stars. 
The corresponding information can be found, for instance, in 
\url{https://exoplanets.nasa.gov/discovery/exoplanet-catalog/}, \url{http://exoplanet.eu}. 
The following objects orbiting our stars are two Neptune-like exoplanet 
around HD 42618 and HD 164595, three gas giants 47 UMa b, c and d around 
HD 95128, a gas giant exoplanet 70 Vir b around HD 117176, a gas giant 
exoplanet 16 Cyg B b around HD 186427, and two gas giant exoplanets around 
HD 187123. Since activity phenomena on hosts of exoplanets can play role 
in influence on planetary environments, further investigations of stellar 
activity are of interest.

\section{ACKNOWLEDGMENTS}

This paper includes data collected with the TESS mission, obtained from the MAST data archive at the Space Telescope Science Institute (STScI). Funding for the TESS mission is provided by the NASA Explorer Program. STScI is operated by the Association of Universities for Research in Astronomy, Inc., under NASA contract NAS 5--26555. This research has made use of data obtained from the Chandra Source Catalog, provided by the Chandra X-ray Center (CXC) as part of the Chandra Data Archive.

\section{FUNDING}

MMK and BAN acknowledge the support of Ministry of Science and Higher Education of the Russian Federation under the grant 075-15-2020-780.

\section{CONFLICT OF INTEREST}

The authors declare that they have no conflicts of interest.

\bibliographystyle{unsrt}

\end{document}